\documentclass[11pt]{article}
\usepackage{hyperref}
\usepackage{graphicx}
\pdfoutput=1
\begin{document}

\title{Electron Beam Artifacts in\\Liquid-Cell Electron Microscopy}

\author{Joseph M. Grogan$^{1}$, Frances M. Ross$^{2}$, Haim H. Bau$^{1}$ \\
\\\vspace{6pt} $^{1}$Department of Mechanical Engineering and Applied Mechanics, \\ University of Pennsylvania, Philadelphia, PA 19104, USA \\
\\\vspace{6pt} $^{2}$IBM T. J. Watson Research Center,  \\ Yorktown Heights, NY 10598, USA}

\maketitle

\begin{abstract}
Bubbles may form when imaging liquids with {\it in situ} liquid-cell electron microscopy. Fluid dynamics videos show beam-induced bubble nucleation and growth. By examining the bubble formation and growth process, we hope to gain a better understanding of interactions between the electron beam and liquids.
\end{abstract}

\section{Main Text}

Liquid-cell electron microscopy has recently emerged as a powerful tool for {\it in situ} studies of nanoscale processes taking place in liquid media with the high resolution of the transmission and scanning transmission electron microscope (TEM and STEM). The technique enables observations of nanoparticle formation, agglomeration, and oriented assembly; electrochemical deposition and etching; and biological structures in their native environment. However, the electron beam interacts with the sample and may produce artifacts that affect the phenomena under study. To utilize the full potential of liquid-cell electron microscopy, it is necessary to obtain a good understanding of the interactions of the electron beam with the imaged medium so that one can mitigate electron beam artifacts, distinguish between beam effects and the phenomena under study, and, in some cases, utilize beam-medium interactions to an advantage.

We combine experiments and theory to investigate and quantify electron beam effects. In particular, we focus on bubble formation and growth. We have encountered two types of bubbles: large bubbles (more common) that spontaneously appear and fill nearly the entire space under the imaging window, and small bubbles (less common) that nucleate on the surface of the membrane window, grow, detach, and migrate away. By quantifying the amount of energy deposited in the sample, we found that under typical operating conditions heating is not significant, but radiolysis can play a key role. We developed a reaction-diffusion model to predict the concentration field of radiolysis byproducts $e_{h}$ (hydrated electron), $H_{3}O^{+}$, $H$, $OH$, $H_{2}$, and $H_{2}O_{2}$ that evolve during irradiation and find that supersaturation of radiolytic $H_{2}$ is a plausible explanation for bubble formation. Linear growth of the bubble radius observed in the experiment is consistent with growth of a spherical bubble fed by a 3D diffusion field with fixed concentration at some distance away. The fixed concentration is due to the steady-state saturation/equilibration of the radiolysis process under constant irradiation. The presence of radiolysis products explains many of the observations of crystal formation, growth, and dissolution recently reported in the literature and observed in our laboratory. Our results suggest approaches to suppress or exploit beam effects, enable improved design of experiments, and facilitate better interpretation of observations resulting from liquid-cell electron microscopy.

\vspace{10 mm}

{\em Acknowledgement:}
This work was supported, in part, by the National Science Foundation grant \# 1066573 and \# 1129722 

\end{document}